\documentclass[aps,12pt,axodraw]{article}
\usepackage{epsfig,sint,macros}

\begin{document}

\begin{titlepage}
\begin{flushright}
MITP/13-044
\end{flushright}

\renewcommand{\thefootnote}{\fnsymbol{footnote}}

\vskip 0.5 cm
\begin{center}
  {\Large\bf 
Reducing the number of counterterms with new minimally doubled actions
\\[0.5ex]}
\end{center}
\vskip 0.5 cm
\begin{center}
{\large Stefano Capitani\footnote{capitan@kph.uni-mainz.de}}
\vskip 0.5cm
Institut f\"ur Kernphysik and HIM (Helmholtz-Institut Mainz), \\
University of Mainz, Johann-Joachim-Becher-Weg 45, D-55099 Mainz, Germany
\vskip 1.0cm
{\bf Abstract}
\vskip 0.35ex
\end{center}

\renewcommand{\thefootnote}{\arabic{footnote}}

\noindent
We study a class of nearest-neighbor minimally doubled actions which depend 
on 2 continuous parameters. We calculate the contributions of the 3 possible
counterterms in perturbation theory, and we find that for each counterterm
there are curves in the parameter space on which its coefficient vanishes.
One can thus construct renormalized actions that contain only 2 counterterms,
instead of the 3 of the standard Karsten-Wilczek or Bori\c{c}i-Creutz actions.

Our investigations suggest the usefulness of analogous nonperturbative
searches for values of the parameters for which the number of counterterms
can be reduced. They can also be an inspiration to undertake a search for 
ultralocal minimally doubled actions with even better counterterm-reducing
properties, including the optimal case in which all counterterms can be
removed. Simulations of the latter actions will be much cheaper than the cases
in which one needs to add counterterms to the bare actions, like the already
conveniently inexpensive standard Karsten-Wilczek fermions.

Finally, we also introduce minimally doubled fermions with 
next-to-nearest-neighbor interactions, which depend on 4 continuous parameters, 
as a further possibility in the search for renormalized actions with no 
counterterms.
\vfill

\begin{center}
July 2013
\end{center}

\eject

\vfill
\eject

\end{titlepage}

\setcounter{footnote}{0}

\section{Introduction}
\label{sec:intro}

Minimally doubled fermions have attracted renewed interest in recent years
as simple lattice formulations which enable Monte Carlo simulations preserving
chiral symmetry for any finite lattice spacing $a$. The Nielsen-Ninomiya theorem
tells us that using two fermion flavors one can maintain an exact continuous
chiral symmetry (of the standard type, i.e. not Ginsparg-Wilson) together with
other convenient field-theoretical properties like locality and unitarity.
Such chiral fermionic formulations can still be kept ultralocal, like
Wilson fermions, but at variance with the latter no tuning of masses is
required, since the continuous chiral symmetry protects masses from additive
renormalization.

Two particular realizations of minimally doubled fermions have been studied 
in some detail in the last few years, the Karsten-Wilczek 
\cite{Karsten:1981gd,Wilczek:1987kw} and Bori\c{c}i-Creutz 
\cite{Creutz:2007af,Borici:2007kz,Creutz:2008sr,Borici:2008ym} actions. 
Since they contain only nearest-neighbor interactions, they are cheap and easy
to simulate. Moreover, they allow the construction of conserved axial currents 
which have a simple expression (again only involving nearest-neighbor sites).

It turns out that for the massless Karsten-Wilczek and Bori\c{c}i-Creutz 
formulations 3 counterterms need to be added to the bare action in order to
remove hypercubic-breaking contributions and so obtain a consistent
renormalized theory
\cite{Capitani:2009yn,Capitani:2009ty,Capitani:2010nn,Capitani:2010ht}. The
coefficients of these counterterms can be determined by enforcing appropriate
conditions from numerical simulations, and once their values are known one can
carry out Monte Carlo computations of physical observables almost almost 
as easily and inexpensively as with Wilson fermions. 

It would be of substantial interest to find minimally doubled actions that
have the correct continuum limit (like the above two cases) but require fewer
counterterms, or even none at all. In the present work we have carried out some
investigations to explore these issues. We only consider orthogonal lattices
and actions which have the correct small $p$ behavior near the two zeros,
that is a leading linear behavior with opposite signs near the two zeros
to give the expected opposite chiralities to the two particles.

The actions that we use here can be seen as generalizations of the
Karsten-Wilczek action, where the distance $2\alpha$ (modulo $2\pi$)
in momentum space between the 2 poles of the quark propagator can be varied
at will, like in the actions proposed in \cite{Creutz:2010qm,Creutz:2011hy}.
A second parameter, $\lambda$, is also introduced, which tunes the $1/a$ mass
of the 14 doublers removed from the naive fermionic action.

It is important to stress that the 3 possible counterterms for all generalized
actions that we introduce here are the same of the standard Karsten-Wilczek
action. This happens because both poles of the quark propagator still lie
entirely on the temporal axis, and thus the temporal direction is always
selected as the special one (irrespective of the values of $\alpha$ and
$\lambda$), and the spinorial structure of all these actions is also the same.
Thus, $P$ is a symmetry, and also $CT$ \cite{Bedaque:2008xs}, but $T$ and $C$
separately are violated (unless the actions are properly renormalized). 

The values of the coefficients of the counterterms for which one is able to 
obtain a consistent renormalized theory depend instead on the particular
choices of $\alpha$ and $\lambda$. 

In the present study we investigate the effects of varying these parameters
to see if one can reduce the contributions due to the counterterms, and
eliminate some of them. In general terms it can be in any case convenient
to have minimally doubled actions where the distance between the two poles
of the quark propagator can be arbitrarily varied, and special values
of this distance could also provide actions more advantageous for simulations
(in that for instance they minimize some artefacts peculiar to these actions).

The values of the coefficients of the counterterms for which the hypercubic
symmetry is restored are continuous real functions of the parameters describing
the actions, $\alpha$ and $\lambda$, and in general there will be values of 
these parameters for which some of these functions vanish. In the case where all
functions happen to become zero for the same values of $\alpha$ and $\lambda$,
no counterterms at all will be required, and one will be able to carry out
consistent simulations using just the tree level actions. This will be then
much cheaper than the already convenient case of (say) Karsten-Wilczek fermions.

We indeed find that for special values of the parameters some of the
3 counterterms can be removed. In perturbation theory one has to compute
the quark self-energy and vacuum polarization and check what kind of 
hypercubic-breaking factors arise from these diagrams. Counterterms need 
indeed to be introduced exactly to act as compensations for these extra 
hypercubic-breaking factors. The main objective of this work is then to see
under which conditions the coefficient of these counterterms can be made to 
vanish. This means that the corresponding functional forms disappear in the
1-loop radiative corrections when special values of $\alpha$ and $\lambda$
are employed.

One of the motivations for our present study has been that, as discovered
in \cite{Capitani:2009yn,Capitani:2009ty,Capitani:2010nn}, for standard 
Bori\c{c}i-Creutz and Karsten-Wilczek fermions the two diagrams of the 1-loop
quark self-energy (sunset and tadpole) always give contributions of opposite
sign to the dimension-three counterterm (which scales as $1/a$). One could then
suspect that using a generalization of these actions an exact cancellation can
occur for some values of the parameters $\alpha$ and $\lambda$, with the effect
that this counterterm (and possibly in general other counterterms) can be
removed from the game. This is indeed what happens.

The main result of this work is that there are a few curves in the parameter 
space spanned by $\alpha$ and $\lambda$ for which one of the counterterms 
can be removed. Thus, the renormalized actions corresponding to these 
particular choices of the parameters require only 2 counterterms. 
There are also values of the parameters for which the renormalized actions have 
no counterterm of order $1/a$, although this occurs in a region of parameter
space where other doublers can appear. Restricting to quenched QCD, we find
that there are many choices of $\alpha$ and $\lambda$ for which only one
counterterm remains.

Even in cases (like the present study) of actions for which it has not been 
possible to remove all counterterms, it is always useful to be able to
accomplish a reduction in the dimensionality of the parameter space of their
coefficients, because it makes their numerical determination easier. 
In particular, if there is only one counterterm left, it is much simpler 
to carry out the determination of its coefficient, because one has to deal 
with just a one-dimensional space instead of a multi-parameter one.

It is also worth remarking that the conserved axial current associated with 
the class of actions introduced here has still the significant advantage of 
being a point-split current involving only nearest-neighbors 
(see Sect.~\ref{sec:conscurr} for its explicit expression).

This article is organized as follows. After presenting in 
Sect.~\ref{sec:actions} the family of nearest-neighbor actions that we 
investigate, together with the Feynman rules needed for our 1-loop calculations,
in Sect.~\ref{sec:conscurr} we give the expressions of the conserved vector
and axial currents. Then we discuss in Sect.~\ref{sec:1-loop} the 1-loop
results for the counterterms, before introducing in Sect.~\ref{sec:next}
minimally doubled fermions with next-to-nearest-neighbor interactions.
Finally, in Sect.~\ref{sec:concl} we present our conclusions together 
with some perspectives.

\section{Nearest-neighbor minimally doubled actions}
\label{sec:actions}

We study the class of bare fermionic actions
\begin{eqnarray}
S^f (x;\alpha,\lambda) &=& a^4 \sum_{x} \bigg[ \frac{1}{2a} \sum_{\mu=1}^4 \Big[
    \overline{\psi} (x) \, 
(\gamma_\mu -i\gamma_4 \,(\lambda +\delta_{\mu 4}(\cot \alpha-\lambda)) )\,
    U_\mu (x) \, \psi (x + a\widehat{\mu}) \nonumber \\
&& \qquad \qquad \qquad -\overline{\psi} (x + a\widehat{\mu}) \,
(\gamma_\mu +i\gamma_4 \,(\lambda +\delta_{\mu 4}(\cot \alpha-\lambda)) )\,
  U_\mu^\dagger (x) \, \psi (x) \Big] \nonumber \\
&& \qquad \quad + \overline{\psi}(x) \, \Big(m_0+\frac{i\gamma_4}{a}\,
  (3\lambda + \cot \alpha)\Big) \, \psi (x) \bigg], 
\label{eq:action}
\end{eqnarray}
in which the interactions are only between nearest-neighbor lattice sites.
These minimally doubled actions, which have as a special direction the temporal
axis (like in the standard Karsten-Wilczek action), satisfy
$\gamma_5$-hermiticity and can be also expressed in the simple form
\begin{equation}
S^f (x;\alpha,\lambda) = a^4 \sum_{x} \overline{\psi} (x) \,\Bigg\{ 
\frac{1}{2} \sum_{\mu} \Bigg[ \gamma_\mu (\nabla_\mu+\nabla^\star_\mu) 
- i a \gamma_4 \,(\lambda +\delta_{\mu 4}(\cot \alpha-\lambda)) \,
\nabla^\star_\mu \nabla_\mu \Bigg] + m_0  \Bigg\} \,\psi (x) ,
\end{equation}
where
\begin{eqnarray}
\nabla_\mu\,\psi (x) &=& 
\frac{U_\mu(x) \psi(x+a\widehat{\mu}) - \psi (x)}{a} 
\label{eq:fcd} \\
\nabla^\star_\mu\,\psi (x) &=& 
\frac{\psi (x) - U_\mu^\dagger(x-a\widehat{\mu}) \psi(x-a\widehat{\mu})}{a}
\label{eq:bcd} 
\end{eqnarray}
are lattice discretizations of the covariant derivative.

In momentum space the Dirac operators of the above minimally doubled fermions 
read, in the free case,
\begin{equation}
  {\cal{D}}^f (p;\alpha,\lambda) = \frac{i}{a} \sum_{\mu=1}^4 \gamma_\mu
   \sin ap_\mu +\frac{i\gamma_4}{a} \,
\Big[ \lambda \sum_{k=1}^3 (1 -\cos ap_k) + \cot \alpha\,(1 -\cos ap_4) \Big] 
+ m_0 .
\label{eq:mom-action}
\end{equation}
The two zeros of ${\cal{D}}^f (p;\alpha,\lambda)$ are located at 
$a\bar{p}_1=(0,0,0,0)$ and $a\bar{p}_2=(0,0,0,-2\alpha)$, and they describe 
two fermions of equal mass and opposite chirality, as the behavior of the 
actions in the vicinity of two zeros, 
$i(\slash{p}-\slash{\bar{p}}_1)~~(=i\slash{p})$ and 
$-i(\slash{p}-\slash{\bar{p}}_2)$, confirms. 

For $\alpha = 0$ and $\alpha = \pi$ the actions become singular 
($\cot \alpha = \infty$), and the range of $\alpha$ can be taken as 
$0 < \alpha < \pi$. Note that, although for the quark propagators corresponding
to $\alpha$ and $\pi-\alpha$ the distance between the poles is the same, 
the actions corresponding to these two choices of $\alpha$ are not equivalent
(even for the same value of $\lambda$). That this is not not just a duplication
can also be seen from our one-loop results (see Fig.~1). 

Varying $\lambda$ does not change the location of any of the zeros of the
actions, as this parameter has only the task of decoupling the 14 other
fermions from the naive fermionic action (which corresponds to the first term
in Eq.~(\ref{eq:mom-action})), giving them a mass of order $1/a$. However,
to avoid the appearance of other doublers it must also be
$\lambda > (1-\cos \alpha)/(2\sin \alpha)$.

It is very important to stress that all the actions presented and considered
in this article have, irrespective of the values of $\alpha$ and $\lambda$,
the correct leading behavior for small $p$. We also remark again that all these 
actions still contain only nearest-neighbor interactions, that is they are 
Wilson-like with hopping terms of only one unit of lattice spacing. For 
this reason they are rather cheap to simulate, and they are a little more
expensive than Wilson fermions because the spinor matrices are slightly more
complicated. The computational effort will be about a few times the one
required for Wilson fermions \cite{numerical}. 

We also note that for $\lambda=1/\sin \alpha$ our actions (\ref{eq:action})
can be cast, after a redefinition of $p_4$, into the actions written by Creutz
in Fourier space in \cite{Creutz:2010qm}. Furthermore, when this choice of
$\lambda$ is taken, the standard Karsten-Wilczek action can be then obtained
as a special case by putting $\alpha=\pi/2$ (which is also the same
as saying $\lambda=1$).

The quark propagator in momentum space corresponding to the minimally 
doubled fermions described by Eq.~(\ref{eq:action}) can be derived by inverting
the Dirac operator of Eq.~(\ref{eq:mom-action}), and is
\begin{equation}
S(p;\alpha,\lambda) = 
    a\,\frac{\displaystyle -i\sum_{\mu=1}^4 \gamma_\mu \sin ap_\mu
    -2i\gamma_4 
    \,\sum_{\mu=1}^4 (\lambda +\delta_{\mu 4}(\cot \alpha-\lambda))\,
    \sin^2 \frac{ap_\mu}{2} +am_0}{D_S(p)} ,
\label{eq:propFP0}
\end{equation}
where the denominator is given by the function
\begin{eqnarray}
 D_S (p;\alpha,\lambda) &=&  \sum_{\mu=1}^4 \sin^2 ap_\mu + 4 \sin ap_4 \,
 \sum_{\mu=1}^4 (\lambda +\delta_{\mu 4}(\cot \alpha-\lambda))\,\sin^2 ap_\mu/2 \\
   & &  +4 \sum_{\mu,\nu=1}^4 
    (\lambda +\delta_{\mu 4}(\cot \alpha-\lambda)) 
    (\lambda +\delta_{\nu 4}(\cot \alpha-\lambda))
    \,\sin^2 \frac{ap_\mu}{2} \,\sin^2 \frac{ap_\nu}{2} 
      +(am_0)^2. \nonumber
\end{eqnarray}
The fermionic vertices which are needed for our 1-loop calculations can also
be obtained easily. The vertex for a quark line which emits a gluon is
\begin{equation}
V_1(p_1,p_2;\alpha,\lambda) = 
   - i g_0 \left(\gamma_\mu \cos \frac{a(p_1+p_2)_\mu}{2}
 +\gamma_4\,(\lambda +\delta_{\mu 4}(\cot \alpha-\lambda))\,
     \sin \frac{a(p_1+p_2)_\mu}{2} \right) ,
\end{equation}
where $p_1$ and $p_2$ are the incoming and outgoing quark momenta, and
for a quark line which emits and then reabsorbs the same gluon is
\begin{equation}
V_2(p_1,p_2;\alpha,\lambda) = \frac{1}{2} i a g_0^2 \left(
    \gamma_\mu \sin \frac{a(p_1+p_2)_\mu}{2} 
 -\gamma_4\,(\lambda +\delta_{\mu 4}(\cot \alpha-\lambda))\,
       \cos \frac{a(p_1+p_2)_\mu}{2} \right) .
\label{eq:v2kw}
\end{equation}
Note that while the quark propagator is invariant under 
$p_k \to -p_k~(k=1,2,3)$, it is not under $p_4 \to -p_4$. This is in accordance
with the fact that $P$ is a symmetry but $T$ is violated. 

As we have already remarked in the Introduction, since the symmetries are
the same of the standard Karsten-Wilczek fermions, the counterterms that
must be added to these generalized actions are the same needed for the
standard Karsten-Wilczek action. In massless quenched QCD only 2 of them are
needed: the fermionic counterterm of dimension four of the form
$\overline{\psi}\,\gamma_4 D_4 \psi$, for which a convenient discretization is
\begin{equation}
d_4 (g_0) \, \frac{1}{2a} \Big( \overline{\psi}(x)
\, \gamma_4 \, U_4 (x) \, \psi (x + a\widehat{4}) 
-\overline{\psi} (x + a\widehat{4}) \, \gamma_4 \,
U_4^\dagger (x) \, \psi (x) \Big) ,
\end{equation}
and the counterterm of dimension three, 
\begin{equation}
\frac{id_3 (g_0)}{a} \,\overline{\psi} (x) \,\gamma_4 \,\psi (x) .
\end{equation}
In full QCD the gluonic part of the action can generate through internal
quark loops a third counterterm, a purely gluonic one, which in continuum form
can be written as
\begin{equation}
\sum_{\rho,\lambda} \Tr\, F_{\rho\lambda}(x) \, F_{\rho\lambda}(x) \,
\delta_{\rho 4} .
\label{eq:ctkwplaq}
\end{equation}
A lattice representation of this counterterm is given, using Wilson's 
plaquette, by
\begin{equation}
d_{\rm{P}}(g_0) \,\frac{N_c}{g_0^2}\, \sum_{\rho,\lambda}
\left( 1-\frac{1}{N_{\rm{C}}}\,\Tr\,P_{4 \lambda}(x) \right) .
\end{equation}

\section{Conserved currents}
\label{sec:conscurr}

The possibility of constructing a conserved axial current, which also
has a simple form and is cheap to use in Monte Carlo simulations, 
constitutes one of the major advantages of using these formulations of 
minimally doubled fermions.

One can derive the expressions of the conserved vector and axial currents 
associated with these fermionic discretizations, by constructing the Ward 
Identities corresponding to the standard infinitesimal chiral transformations
(for instance along the lines of \cite{Bochicchio:1985xa}). The expression
of the conserved axial current that one obtains in this way is
\begin{eqnarray}
A_\mu^{\mathrm cons} (x;\alpha,\lambda) &=& \frac{1}{2} \bigg(
   \overline{\psi} (x) \, (\gamma_\mu -
      i\gamma_4 \, (\lambda+\delta_{\mu 4}(\cot \alpha-\lambda)) ) \,
   \gamma_5 \, U_\mu (x) \, 
   \psi (x+a\widehat{\mu}) 
  \label{eq:noether-axial} \\ 
 && \qquad \quad + \overline{\psi} (x+a\widehat{\mu}) \, 
    (\gamma_\mu +
      i\gamma_4 \, (\lambda+\delta_{\mu 4}(\cot \alpha-\lambda)) ) \,
  \gamma_5 \, U_\mu^\dagger (x) \, \psi (x) \bigg) \nonumber \\
 && + \frac{d_4 (g_0)}{2} \bigg(
   \overline{\psi} (x) \, \gamma_4 \gamma_5 \, U_4 (x) \, 
   \psi (x+a\widehat{4}) 
  + \overline{\psi} (x+a\widehat{4}) \, \gamma_4 \gamma_5 
   \, U_4^\dagger (x) \, \psi (x) \bigg) . \nonumber
\end{eqnarray}
This is an exactly conserved quantity for any finite value of the lattice 
spacing $a$, and only involves nearest-neighbor sites. This is particularly
important, as not many fermionic formulations exist for which a conserved
axial current exists and is of such a simple form. Note also that only the
quark counterterm of dimension 4 enters in the explicit expression of the
renormalized conserved current. 

Finally, to obtain the conserved vector current corresponding to our class
of actions it is sufficient to simply remove the $\gamma_5$ matrices 
from the above expression.

\section{1-loop curves of zeros} 
\label{sec:1-loop}

The coefficients of the counterterms for which these actions are properly
renormalized can be determined by computing in the cases of $d_3$ and $d_4$
the quark self-energy. Their values are the ones for which the
hypercubic-breaking factors in the radiative corrections disappear.
In the case of $d_{\rm{P}}$ one enforces the restoration of the hypercubic
symmetry on the renormalized vacuum polarization of the gluon,
$\Pi_{\mu\nu}(p)$. For these calculations we have used Wilson's plaquette action
in a general covariant gauge (where $\partial_\mu A_\mu=0$), and also put
$m_0=0$.

Due to the non-trivial form of the denominator of the quark propagator, 
it is not possible to provide results with an analytic dependence on
$\alpha$ or $\lambda$. The search for the special values of these parameters 
which remove the hypercubic-breaking factors in the 1-loop quark self-energy 
and vacuum polarization must then be carried out numerically.

The tadpole of the self-energy however can be calculated analytically, and
its result has a simple dependence on $\alpha$ and $\lambda$. In a general
covariant gauge one obtains\,\footnote{The quantity $Z_0$ is an often-recurring
lattice integral, defined as
\cite{GonzalezArroyo:1981ce,Ellis:1983af,Capitani:2002mp} 
\begin{equation}
  Z_0 = \int_{-\pi/a}^{\pi/a} \frac{d^4p}{(2\pi)^4}
       \frac{1}{\widehat{p}^2} = 0.1549333\ldots 
       = \frac{24.466100\ldots}{16\pi^2},\qquad
       \widehat{p}^2=\frac{4}{a^2}\,\sum_\mu \sin^2\Big(\frac{ap_\mu}{2}\Big) .
\end{equation}
}
\begin{eqnarray}
T & = & \frac{1}{a^2} \cdot \frac{Z_0}{2} \Big(1-\frac{1}{4}(1-\xi)\Big)
\cdot i a g_0^2 \,C_F \sum_{\mu=1}^4 \Big( \gamma_\mu a p_\mu
 -\gamma_4\,(\lambda+\delta_{\mu 4}(\cot \alpha-\lambda)) 
(1+O(a^2) \Big)  \nonumber \\
& = & g_0^2 \, C_F \, \frac{Z_0}{2} \Big(1-\frac{1}{4}(1-\xi)\Big) 
    \,\Big( i\slash{p}
  - \frac{i\gamma_4}{a}\,(3\lambda+\cot \alpha) \Big) +O(a) ,
\label{eq:tadpole}
\end{eqnarray}
where terms of $O(a)$ and higher are eventually set to zero. Notice that
for $\lambda=1$ and $\alpha=\pi/2$ we recover the result of the standard
Karsten-Wilczek action, which was given in
\cite{Capitani:2009ty,Capitani:2010nn,Capitani:2010ht}. 

The result for the $\,i\slash{p}\,$ term is the same of Wilson fermions. 
The other term, which is linearly divergent as $1/a$, has a functional form 
already present in the bare minimally doubled action (\ref{eq:action}),
where however its coefficient is a fixed number. In the renormalized action
instead it becomes a counterterm, whose coefficient must be properly adjusted
as a function of the gauge coupling.

One important feature of the $1/a$ term in the result of the tadpole $T$ 
is that not only it diverges when $a\to 0$ (contributing to the relevant
counterterm $d_3$), but also diverges when $\alpha \to 0$, in the latter case
with a behavior which goes like $1/\sin \alpha$ (for fixed lattice spacing).
It also diverges at the other end of the validity range, $\alpha \to \pi$,
with a similar behavior.

To carry out the calculations of the two other diagrams required for 
the tuning of the counterterms (the sunset of the quark self-energy,
and the quark loop of the vacuum polarization of the gluon), we have used 
a set of computer codes written in the algebraic manipulation language FORM 
\cite{Vermaseren:2000nd,Vermaseren:2008kw}, extended to include the special
features of the actions presented here. We have checked that the results 
are gauge invariant, in particular that the part which should vanish in 
Feynman gauge, which is proportional to $(1-\xi)$, is indeed zero 
(within integration errors) in the results of $d_3$ and $d_4$, for several
choices of $\alpha$ and $\lambda$. For the polarization of the vacuum, which at 
one loop does not involve internal gluon propagators, this kind of check is 
not possible. In the latter case we have then also checked that, apart from
recovering the standard Karsten-Wilczek result as a special case, quantities
which are expected to vanish, like for example the $p_1p_2$ contribution to
$\Pi_{11}(p)$, are indeed zero within a large numerical precision (for several
choices of $\alpha$ and $\lambda$), in spite of the fact that the corresponding
integrand functions (as FORM outputs) are still very large expressions
containing many thousands of terms.

Our main results are summarized in Fig.~1, which shows the curves for which
each counterterm has a vanishing coefficient. Some of the points that we have
obtained are also given explicitly in Table 1. Our calculations show that there
are no intersections between these curves of zeros. The curve corresponding to
a zero of $d_4$ (the fermionic counterterm of dimension four) extends over all
possible range of $\alpha$, and it is a convex function whose minimum is not
at $\pi/2$, but still not too distant from it. We can see that this curve
is not symmetric with respect to the reflection $\alpha \to \pi/2 -\alpha$,
and this is consistent with the fact that, although the distance between the two
poles of the quark propagator does not change when $\alpha \to \pi/2 -\alpha$,
these two choices of $\alpha$ do not correspond to equivalent actions. 

We also notice that the curve corresponding to the vanishing of the 
dimension-3 counterterm, $d_3$, has instead a domain which is restricted to 
$\alpha > \pi/2$. For $\alpha \to \pi/2$ (from above) along this curve,
$\lambda$ goes asymptotically to zero, while for $\alpha \to \pi$ what happens
is that $\lambda$ begins to grow very rapidly. This is a behavior which is
substantially different from the one of the fermionic dimension-4 counterterm.

Notice also that there are 2 curves of zeros for the gluonic counterterm,
$d_{\rm{P}}$, whereas there is one only curve of zeros for each of the 2
fermionic counterterms (the only ones which are relevant in the quenched case).

In Fig.~1 we only show results for $\lambda < 3$, since the trend looks clear.
From further investigations it seems that there are no intersections even when
one goes to larger values of $\lambda$. When $\lambda$ becomes very large, the
curves of zeros get closer and closer to $\alpha = 0$ or $\alpha = \pi$, and
at the same time the distance between the two poles of the quark propagator
becomes smaller and smaller. Eventually at these limiting values of $\alpha$
the action becomes singular (as $\cot \alpha = \infty$). That for very large
values of $\lambda$ one gets near a singular situation can also be seen from
the fact that the tadpole $T$ diverges when $\alpha \to 0$ or $\alpha \to \pi$
(see Eq.~(\ref{eq:tadpole})). We also find that in this region the errors on the
numerical results increase substantially, and the zeros of the coefficients of
the counterterms cannot be resolved with a good precision.

At the end one has to consider the region in parameter space corresponding
to $\lambda > (1-\cos \alpha)/(2\sin \alpha)$, because otherwise additional
doublers can be generated and then obviously we are not anymore in a situation
of minimally doubled fermions. In our 1-loop calculations we have observed that
as $\lambda$ gets smaller the convergence of the integrals for the vacuum
polarization diagram rapidly worsens, and it becomes difficult to compute these
integrals with a good precision. This is why we have not completed the curves
for $d_{\rm{P}}$ towards $\lambda=0$. This is also in part due to the numerous
integrals required for this diagram, which arise from a quadratic Taylor
expansion in the lattice spacing and are thus quite expensive to evaluate.
Since the number of terms is some orders of magnitude larger than in the case
of the quark self-energy, the search for the zeros of $d_{\rm{P}}$ turns out to
be much more expensive than for the fermionic counterterms $d_3$ and $d_4$, and
the precision that can be accomplished is much smaller. For the fermionic
counterterms the zeros can instead be easily determined with a much lower error.

Of course one can always compute a few selected zeros in a small region 
of the space of parameters with very high precision, if needed. The complete
mapping of the whole space of parameters would require however an extremely
larger computational effort, so that at the end only a much lower precision
can be accomplished. 

At any rate, the main purpose of the present investigations is not the
computation of all zeros with a high precision, but we want rather to show
that such curves of zeros exist and see what shape they have.
There is no need at this stage to determine these curves of zeros with high 
precision, although of course when one will eventually be able to construct a 
nonperturbatively renormalized action which needs just one (or no) counterterm, 
a determination with higher precision of the corresponding points
will be (at that point) desirable.

As was reported in \cite{Capitani:2010nn}, in the special case $\alpha=\pi/2$
there is no way for $d_3$ (the coefficient of the $O(1/a)$ counterterm) 
to become zero by adjusting the value of $\lambda$. The reason is that, 
although the tadpole and sunset diagrams have opposite sign and their absolute 
values decrease as $\lambda \to 0$, the sunset always remains in magnitude
much smaller than the tadpole, and so a cancellation is never possible. 
The calculations presented in this work show that such a cancellation can take
place when $\alpha > \pi/2$ (as can be checked in Fig.~1). Indeed, for any
$\pi/2 < \alpha < \pi$ there is always a value of $\lambda$ for which a
cancellation occurs, but not for $\alpha = \pi/2$ (the standard Karsten-Wilczek
action). This is an example of the usefulness of having actions in which
the distance between the 2 poles of the quark propagator can be varied.

As already mentioned in Sect.~\ref{sec:actions}, in the particular case 
$\lambda=1/\sin \alpha$ our actions (\ref{eq:mom-action}) can be cast, after
a redefinition of $p_4$, into the actions proposed by Creutz in
\cite{Creutz:2010qm}, which in the free massless case read
\begin{equation}
  {\cal{D}}^C (p;\alpha) = \frac{i}{a} \sum_{k=1}^3 \gamma_k \sin ap_k 
  +\frac{i\gamma_4}{a \sin \alpha } \, 
  \Big( \cos \alpha +3 -\sum_{\mu=1}^4 \cos ap_\mu \Big) .
\end{equation}
It is interesting that, as we have also indicated in Fig.~1, there are two
points for which the curve $\lambda=1/\sin \alpha$ intersects the curve of
zeros of $d_4$. As a consequence the action proposed by Creutz, which in
general requires 3 counterterms, needs only 2 of them when either of the
following two choices of $\alpha$ is made: $(\alpha,\lambda)=(1.47,1.01)$ or
$(\alpha,\lambda)=(2.41,1.49)$. In both cases it is the fermionic counterterm
of dimension four which is eliminated.

\section{Next-to-nearest-neighbor minimally doubled actions}
\label{sec:next}

In the search for actions for which intersections between the curves of zeros 
exist (so that 2 or even more of the possible counterterms can then be
removed), one can think of widening the pool by considering also couplings 
between next-to-nearest-neighbor lattice sites. In the quest for minimally 
doubled actions without counterterms, investigating such kind of actions could 
turn out at the end to be rewarding. We do not know in fact whether there 
could be theoretical impediments in principle to countertermless minimally 
doubled actions when one only considers nearest-neighbor interactions. 
It is conceivable that introducing interactions also at distance $2a$ or larger 
could allow actions with different kinds of properties, and the hope is that 
at the end some of these actions will not require any counterterms to be
properly renormalized.

We find then useful to propose here a first example of a class of 
minimally doubled actions with next-to-nearest-neighbor interactions:
\begin{eqnarray}
S^f_{ntn} (x;\alpha,\lambda,\lambda',\rho) 
   &=& a^4 \sum_{x} \bigg[ \frac{1}{2a} \sum_{\mu=1}^4 \Big[
    \overline{\psi} (x) \,(\gamma_\mu -i\gamma_4 \,
    f^{(1)}_\mu) \, U_\mu (x) \, \psi (x + a\widehat{\mu}) \nonumber \\
&& \qquad \qquad \qquad -\overline{\psi} (x + a\widehat{\mu}) \,
   (\gamma_\mu +i\gamma_4 \, f^{(1)}_\mu) \, U_\mu^\dagger (x) \, \psi (x) \Big]
   \nonumber \\
  && \qquad \quad
    + \frac{i}{4a}\sum_{\mu=1}^4 \, f^{(2)}_\mu \, \Big[ \overline{\psi} (x) \,
    \gamma_4 \, U_\mu (x) \, U_\mu (x + a\widehat{\mu}) 
    \, \psi (x + 2a\widehat{\mu}) \nonumber \\
&& \qquad \qquad \qquad \qquad \quad  
  +\overline{\psi} (x + 2a\widehat{\mu}) \, \gamma_4 \,
  U_\mu^\dagger (x + a\widehat{\mu}) \, U_\mu^\dagger (x) 
   \, \psi (x) \Big] \nonumber \\
&& \qquad \quad + \overline{\psi}(x) \,
   \Big(m_0+\frac{i\gamma_4}{a}\, f^{(0)} \Big)\,\psi (x) \bigg] , 
\label{eq:nextaction}
\end{eqnarray}
where 
\begin{eqnarray}
f^{(0)}(\alpha,\lambda,\lambda',\rho) &=& 3\lambda + \frac{9}{2}\lambda' + 
   \Big(\rho+ \frac{3}{4} \frac{1-\rho}{\sin^2 \alpha} \Big)\cot \alpha , \\
f^{(1)}_\mu(\alpha,\lambda,\lambda',\rho) &=& \lambda+2\lambda' 
   +\delta_{\mu 4}\,\Big(\Big(\rho+ \frac{1-\rho}{\sin^2 \alpha}\Big)
   \cot \alpha-\lambda-2\lambda'\Big) , \\
f^{(2)}_\mu(\alpha,\lambda',\rho) \ \  &=& \lambda' +\delta_{\mu 4}\,\Big(
   \frac{1-\rho}{2\sin^2 \alpha} \cot \alpha-\lambda'\Big)
\end{eqnarray}
are functions diagonal in spinor and color space. Note that there are
simple relations between these functions, and if one defines
\begin{equation}
f^{(h)}_\mu(\alpha,\lambda,\rho) = 
\lambda +\delta_{\mu 4}\,\Big(\rho \cot \alpha - \lambda \Big)
\end{equation}
then knowing $f^{(1)}_\mu$ one can obtain
\begin{eqnarray}
f^{(2)}_\mu &=& \frac{1}{2} \, \Big(f^{(1)}_\mu-f^{(h)}_\mu\Big) , \\
f^{(0)} &=& \sum_{\mu=1}^4 \Big(\frac{3}{4}\, f^{(1)}_\mu +f^{(h)}_\mu\Big) 
= \sum_{\mu=1}^4 \Big(f^{(h)}_\mu + \frac{3}{2}\, f^{(2)}_\mu\Big) .
\end{eqnarray}
These actions satisfy $\gamma_5$-hermiticity, and like for the standard
Karsten-Wilczek action the temporal direction is again the special one which
is selected and which then breaks hypercubic symmetry. The symmetries are then
the same of the Karsten-Wilczek action, that is $P$ is conserved, and also
$CT$, but $T$ and $C$ separately are violated unless the action is properly
renormalized. Then, the counterterms that have to be added to these
generalized actions are again the same of the standard Karsten-Wilczek action.

The corresponding momentum-space actions are given in the free case
by\,\footnote{Notice that for $\lambda'=0$ and $\rho=1$ one falls back to the
case of the nearest-neighbor actions (\ref{eq:action}) that we have studied
in the rest of the article.}
\begin{eqnarray}
  {\cal{D}}^f_{ntn} (p;\alpha,\lambda,\lambda',\rho) 
    &=& \frac{i}{a} \sum_{\mu=1}^4 \gamma_\mu
   \sin ap_\mu +\frac{i\gamma_4}{a} \, \Big\{ \sum_{k=1}^3 \Big( 
   \lambda\,(1 -\cos ap_k) + \lambda'\,(1 -\cos ap_k)^2 \Big) 
 \nonumber \\
  && \quad + \cot \alpha \,\Big( \rho\,(1 -\cos ap_4) 
   +\frac{1-\rho}{2\sin^2 \alpha}\,(1 -\cos ap_4)^2 \Big) \Big\} + m_0 .
\label{eq:mom-nextaction} 
\end{eqnarray}
The parameter $\alpha$ regulates the distance between the two zeros, which are 
at the same positions $a\bar{p}_1=(0,0,0,0)$ and $a\bar{p}_2=(0,0,0,-2\alpha)$
as in the nearest-neighbor actions. That there are only two zeros is certainly 
the case if $-3 \le \rho \le 1$ (as we have numerically verified), and 
$-\pi/2 < \alpha < \pi/2$ as before. For choices of $\rho$ outside of this
range, additional zeros can in general appear, and one can still get minimally
doubled actions but only for a restricted domain of $\alpha$ (whose extension
depends on the value of $\rho$).\,\footnote{For example, with the choice
$(\alpha,\rho)=(0.1,1.1)$ and for $\vec{p}=(0,0,0)$, the action is proportional
to $\gamma_4$, and its coefficient a function of $p_4$ only, which has four
intersections with the $p_4=0$ axis. It seems then that obtaining minimally
doubled actions is not a completely trivial business.} Moreover, one must 
also take $\lambda + 2\lambda' >  -\min \, \{\sin x + \cot \alpha \,(
\rho\,(1 -\cos x) +(1-\rho)\,(1 -\cos x)^2/(2\sin^2 \alpha)) \}/2$ to ensure
that there are no more than two fermions.

It is worth noting that the above actions in position space can also be written 
more concisely in the simple form
\begin{eqnarray}
S^f_{ntn} (x;\alpha,\lambda,\lambda',\rho) &=& a^4 \sum_{x} \overline{\psi} (x) 
\,\Bigg\{ \sum_{\mu} \Bigg[ \frac{1}{2} \,\gamma_\mu (\nabla_\mu+\nabla^\star_\mu) 
\label{eq:concisenextaction} \\
&& \qquad \qquad \qquad
  - i a \gamma_4 \Big\{ \frac{1}{2}\, f^{(1)}_\mu \, \nabla^\star_\mu \nabla_\mu
  -f^{(2)}_\mu \, \widetilde{\nabla}^\star_\mu \widetilde{\nabla}_\mu \Big\} \Bigg]
  +m_0 \Bigg\} \,\psi (x), \nonumber 
\end{eqnarray}
where, in addition to the standard $\nabla_\mu$ and $\nabla^\star_\mu$ of 
Eqs.~(\ref{eq:fcd}) and (\ref{eq:bcd}), one has also introduced another
discretization for the lattice covariant derivative, extending this time over
two lattice sites:
\begin{eqnarray}
\widetilde{\nabla}_\mu\,\psi (x) &=& \frac{U_\mu(x) \, U_\mu (x + a\widehat{\mu})
  \,\psi(x+2a\widehat{\mu}) - \psi (x)}{2a}, \\
\widetilde{\nabla}^\star_\mu\,\psi (x) &=& 
\frac{\psi (x) - U_\mu^\dagger(x-a\widehat{\mu}) \,U_\mu^\dagger(x-2a\widehat{\mu})
  \,\psi(x-2a\widehat{\mu})}{2a}.
\end{eqnarray}
Thus, two different discretizations of the same continuum covariant derivative 
appear at the same time in these actions. But the overall structure 
(in particular regarding Laplacians) is very similar to the one of the 
nearest-neighbor actions (the two kinds of lattice derivatives differing only 
by terms of $O(a)$). 

Note also that in the concise notation (\ref{eq:concisenextaction})
it is apparent that there is no mass term left if one sets $m_0=0$ (as also 
was true for the nearest-neighbor actions). The terms proportional to 
$\,i\overline{\psi}(x)\gamma_4\psi(x)/a\,$ in Eqs.~(\ref{eq:action}) and 
(\ref{eq:nextaction}) are in fact part of the various Laplacians.

Our primary motivation for introducing these next-to-nearest-neighbor actions 
is that for special choices of the parameters one could hit on renormalized 
actions which do not require any counterterms. The fact that there are 4 
parameters, and not just 2 as in the nearest-neighbor actions, should result 
in many more curves on which the counterterms become zero and, above all, 
more chances for intersections among these curves. It could then happen that
there exist some values of the parameters for which one ends up with just one 
counterterm, or none at all.\,\footnote{Actually the ``curves'' are here
likely to be 3-dimensional manifolds in the 4-dimensional parameter space,
separating domains in which the coefficients are positive or negative.
It could then happen that there exist renormalized minimally doubled actions
without counterterms, which are described by 1-dimensional curves in the space
of parameters.} 

Of course to explore adequately this larger parameter space will be more 
expensive than for the nearest-neighbor actions.\,\footnote{Although one 
could start with significant subspaces which are cheaper to explore, like
$\lambda=0$ or $\lambda+\lambda'=1$.}
To carry out a complete mapping of the curves of zeros on the whole space 
of parameters, even just in perturbation theory, will require a substantial 
computational effort, which is left for future work. 

It is probably also not too difficult to go one step further and construct
minimally doubled fermions with hopping terms extending to three (or more) 
lattice spacings. This will enlarge even further the space in which to search
for actions which do not require counterterms, although incrementing the 
range of the couplings renders such actions increasingly less convenient
for simulations.

\section{Conclusions} 
\label{sec:concl}

One important result of the present work is that there are curves in the
space spanned by the two parameters $\alpha$ and $\lambda$, which define the 
nearest-neighbor actions, on which some of the counterterms vanish, possibly
because of an increased symmetry of the corresponding renormalized actions. 
Then the counterterms needed for a consistent 1-loop renormalized theory
are fewer than the 3 required for the standard massless Karsten-Wilczek and
Bori\c{c}i-Creutz actions. We find that there are actually many choices of
$\alpha$ and $\lambda$ for which a counterterm can be left out, and this 
renders the actions corresponding to these particular values of $\alpha$
and $\lambda$ rather cheap and convenient to simulate.

If some of the curves of zeros had an intersection point, this would be quite 
advantageous, since the corresponding values of $\alpha$ and $\lambda$ would
provide a renormalized minimally doubled action which needs only one
counterterm. Unfortunately the curves that we have obtained seem not to
intersect, and so one remains always with at least 2 counterterms, at least
in perturbation theory and within the families of actions considered in this
paper. However one can still choose in some convenient way which counterterms
to keep and which one to discard. In the quenched case, one has the possibility
to construct an action with just one counterterm. In full QCD, one can choose
to use only the 2 fermionic counterterms, so that there is then no need to 
fine-tune and employ a gluonic operator of the $FF$ form. 

Although the conclusions presented in this work arise from perturbative
calculations, it is likely that also in numerical simulations the removal
of some counterterms can be accomplished for appropriate choices of the
parameters. Of course the nonperturbative curves of zeros will be (to a
smaller or greater extent) different from the ones presented in this article,
and one should check whether the qualitative pattern of the curves that we
have found is also reproduced nonperturbatively. Since the coefficients of the
counterterms appear to be rather smooth functions of the parameters of the
action (except possibly when one gets close to the limits of the domain of
$\alpha$, i.e. $\alpha \sim 0$ and $\alpha \sim \pi$), it will probably be
not too expensive to first perform a quick rough tuning of the parameters
around the curves of zeros that we have calculated perturbatively, and
subsequently to compute with more precision the positions of these
nonperturbative zeros using a much finer tuning. It could also turn out that
the locations of these zeros do not differ too much from the perturbative
results (at least in some regions of the space of parameters), and so one
could take the perturbative results as a good starting guess.

It is also possible that nonperturbatively the vanishing of the counterterm of
dimension 3 occurs in the region where there is minimal doubling. Since this is 
the only relevant counterterm (it has dimension 3), in this case only 2 
marginal counterterms (of dimension 4, whose coefficient moreover is likely
to be rather small 
\cite{Capitani:2009yn,Capitani:2009ty,Capitani:2010nn,Capitani:2010ht})
would remain to be tuned in order to carry out consistent Monte Carlo
simulations, leading to milder numerical cancellations.

In principle some intersection points could appear at the nonperturbative level.
Therefore, even though we have not been able here to remove all counterterms
(having only used perturbation theory), it could well happen that
nonperturbatively an intersection point does exist. In the case in which the
(nonperturbative) curves indeed intersect, the intersection points will be the
most important numbers to find. Since there will not likely be many of them,
it will not be overly expensive to determine them with high precision.
This would make possible to simulate renormalized minimally doubled actions
with at most one counterterm.

We have seen that in the particular case of the class of actions that we have 
presented here, there are numerous choices of parameters which give just 
2 counterterms (and hence only 1 in the unquenched case). It is possible that 
still cleverer minimally doubled actions can be constructed, which would be
able to accomplish an even greater reduction of the number of counterterms,
including the optimal situation where a maximal reduction can be accomplished,
that is no counterterms at all are needed. In this case Monte Carlo simulations 
employing just the bare action will be sufficient for the extraction of 
significant physical results, and no tuning of counterterms will be required
to simulate such an action. 

In any case, leaving aside the removal of counterterms, it is always useful
to possess as many different minimally doubled actions as possible and keep on
trying to construct new ones, because some particular actions could turn out
to have better theoretical or practical properties. The effective amount of
important physical quantities such as the mass difference between the
$\pi^\pm$ and the $\pi^0$, or of mass splittings within otherwise degenerate
multiplets, could turn out to be rather small for a few of these actions and not
for all other ones. In general terms it can be convenient to have (like here)
minimally doubled actions where the distance between the two poles of the quark
propagator can be arbitrarily varied. Special values of this distance could
also provide actions which could turn out to be more advantageous for Monte
Carlo simulations.\,\footnote{We have tried without success to construct some
Bori\c{c}i-Creutz-like action in which the distance between the zeros can be
arbitrarily chosen and which has the correct continuum limit. However the
Bori\c{c}i-Creutz action appears to be much more constrained in its form than
the Karsten-Wilczek action and to leave little room for changes.}

Thus, this work can also be considered as an inspiration to undertake further
searches for new minimally doubled actions which require a reduced number 
of counterterms, and possibly (in the best of cases) none at all.
While we think reasonable to concentrate most efforts on formulations defined
on orthogonal lattices, one can imagine to loosen the ultralocality
constraint (which was instead retained in all of this work), and then also
include actions containing flavored mass terms, like the ones constructed in 
\cite{Creutz:2010qm,Creutz:2011hy,Creutz:2010bm,Creutz:2011cd}, in which
the single flavors can be separately identified. Further theoretical
investigations, and other recipes for constructing minimally doubled actions
like the twisted-ordering method presented in 
\cite{Creutz:2010cz,Misumi:2010ea,Kimura:2011ik} (see also the overview in
\cite{Misumi:2012eh}), could suggest how to steer these searches also in
new directions. For every new candidate action, one of the most important
issues will be to understand what kind of counterterms can appear, and 
how many of them are in principle required.

In any case, the next-to-nearest-neighbor actions (depending on 4 parameters)
that we have introduced in Sect.~\ref{sec:next} could also be taken as a
starting point for a special direction in this undertaking, especially if it
turns out that next-to-nearest-neighbor interactions possses some fundamental 
feature different from the nearest-neighbor case, in particular with respect to 
the type of counterterms which can arise.

\begin{figure}[t]
\begin{center}

\includegraphics[height=14.7cm]{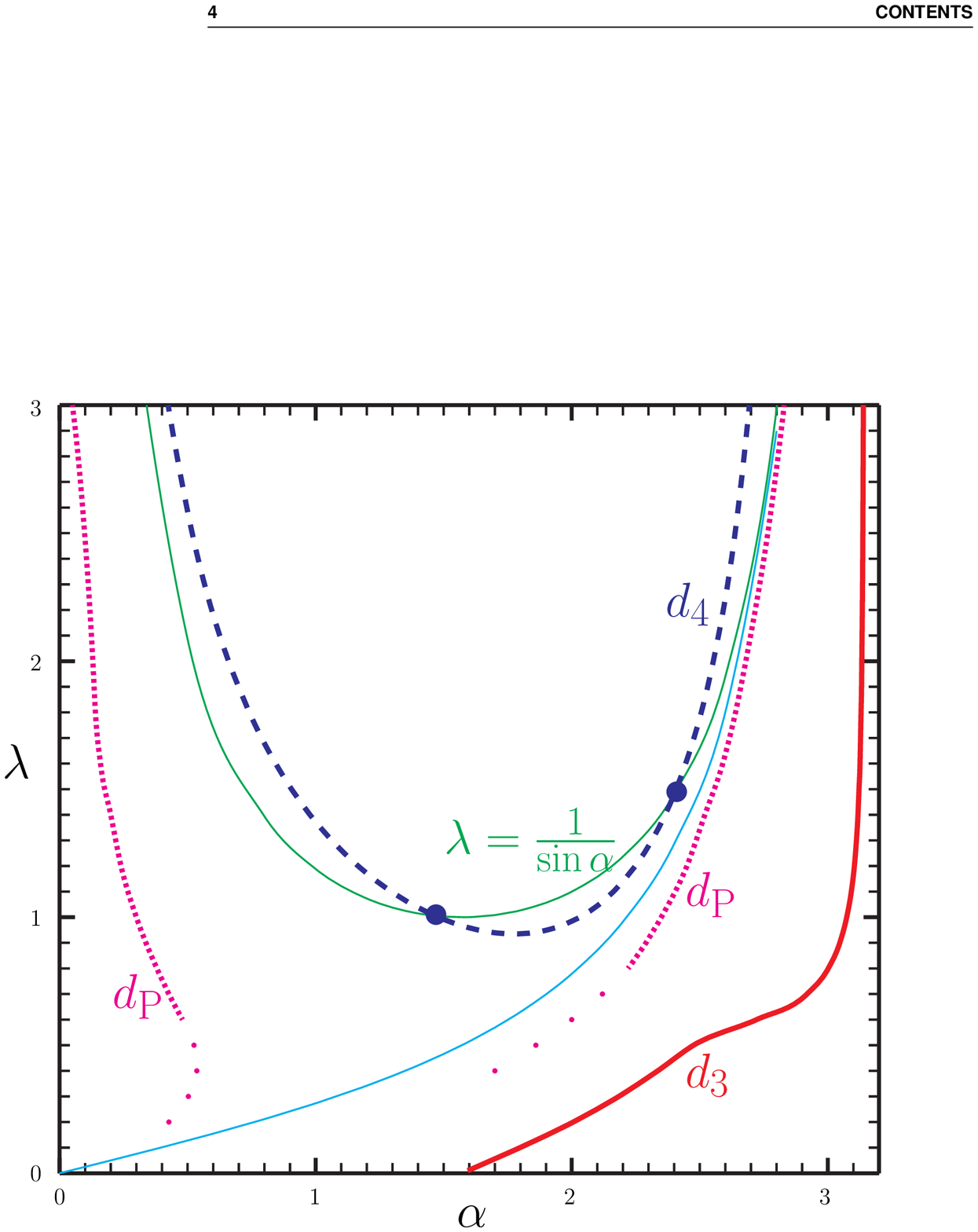}

\end{center}
\caption{\small Curves of zeros for the coefficients of the various 
counterterms. The continuous curve indicates the vanishing of $d_3$, the 
dashed curve the vanishing of $d_4$, and the dotted curves the vanishing of 
$d_{\rm{P}}$. The curves are interpolations of the many points obtained from 
1-loop calculations (some of them are shown in Table 1).
The curve $\lambda=1/\sin \alpha$, which corresponds to the action proposed
by Creutz, it also drawn, showing that for this action two choices of
$\alpha$ exist for which the fermionic counterterm of dimension 4 is not
required.
Shown is also the curve $\lambda > (1-\cos \alpha)/(2\sin \alpha)$, below
which additional doublers can appear.}  
\label{fig:doublers}
\end{figure}

\begin{table*}[htp]
\begin{center}
\caption{\label{tab:atable}
Values of $\alpha$ at which the various counterterms vanish, for selected values
of $\lambda$. The errors in the case of $d_{\rm{P}}$ are much larger, 
of about $5\cdot 10^{-2}$.}
\vspace{0.4cm}
\begin{tabular}{|l|l|l|l|l|}
\hline
$\lambda$ & $d_3$ & $d_4$ & $d_{\rm{P}}$ & $d_{\rm{P}}$ 
\\ \hline 
  0.6   & 2.722   &                 & 0.48   &       \\
  0.7   & 2.919   &                 &        &       \\
  0.8   & 3.002   &                 & 0.38   & 2.22  \\
  0.9   & 3.047   &                 &        &       \\
  0.94  &         & 1.694           &        &       \\
  0.945 &         & 1.661;~~1.882   &        &       \\
  0.95  &         & 1.635;~~1.904   &        &       \\
  0.96  &         & 1.594;~~1.940   &        &       \\
  0.97  &         & 1.559;~~1.968   &        &       \\
  1.0   & 3.074   & 1.478;~~2.032   & 0.30   & 2.35  \\
  1.05  &         & 1.376;~~2.107   &        &       \\
  1.1   & 3.092   & 1.295;~~2.163   &        &       \\
  1.15  &         & 1.226;~~2.209   &        &       \\
  1.2   & 3.104   & 1.166;~~2.248   & 0.24   & 2.45  \\
  1.25  &         & 1.112;~~2.283   &        &       \\
  1.3   & 3.112   & 1.063;~~2.313   &        &       \\
  1.35  &         & 1.019;~~2.340   &        &       \\
  1.4   & 3.118   & 0.978;~~2.365   & 0.20   & 2.52  \\
  1.45  &         & 0.941;~~2.388   &        &       \\
  1.5   & 3.122   & 0.906;~~2.408   &        &       \\
  1.55  &         & 0.874;~~2.427   &        &       \\
  1.6   & 3.126   & 0.844;~~2.445   & 0.16   & 2.59  \\
  1.7   & 3.128   & 0.789;~~2.477   &        &       \\
  1.8   & 3.130   & 0.741;~~2.506   & 0.14   & 2.64  \\
  1.9   & 3.132   & 0.699;~~2.531   &        &       \\
  2.0   & 3.133   & 0.660;~~2.553   & 0.13   & 2.68  \\
  2.5   & 3.137   & 0.518;~~2.638   &        &       \\
  2.6   &         &                 & 0.09   & 2.78  \\
  3.0   & 3.139   & 0.424;~~2.694   & 0.05   & 2.83  \\ \hline    
\end{tabular}
\end{center}
\end{table*}

\end{document}